# Rotational and translational diffusion of anisotropic gold nanoparticles in liquid crystals controlled by varying surface anchoring

Bohdan Senyuk,[1] David Glugla,[2] and Ivan I. Smalyukh[1,2,3,4*]

[1]*Department of Physics, University of Colorado at Boulder, Boulder, Colorado 80309, USA*

[2]*Department of Electrical, Computer, and Energy Engineering, University of Colorado at Boulder, Boulder, Colorado 80309, USA*

[3]*Liquid Crystals Materials Research Center and Materials Science and Engineering Program, University of Colorado at Boulder, Boulder, Colorado 80309, USA*

[4]*Renewable and Sustainable Energy Institute, National Renewable Energy Laboratory and University of Colorado at Boulder, Boulder, Colorado 80309, USA*

*Email: ivan.smalyukh@colorado.edu

**Abstract**

We study translational and rotational diffusion of anisotropic gold nanoparticles (NPs) dispersed in the bulk of a nematic liquid crystal fluid host. Experimental data reveal strong anisotropy of translational diffusion with respect to the uniform far-field director, which is dependent on shape and surface functionalization of colloids as well as on their ground-state alignment. For example, elongated NPs aligned parallel to the far-field director translationally diffuse more rapidly along the director whereas diffusion of NPs oriented normal to the director is faster in the direction perpendicular to it while they are also undergoing elasticity-constrained rotational diffusion. To understand physical origins of these rich diffusion properties of anisotropic nanocolloids in uniaxially anisotropic nematic fluid media, we compare them to diffusion of prolate and oblate ellipsoidal particles in isotropic fluids as well as to diffusion of shape-isotropic particles in nematic fluids. We also show that surface functionalization of NPs with photosensitive azobenzene groups allows for *in situ* control of their diffusivity through *trans*-*cis* isomerization that changes surface anchoring.

## I. INTRODUCTION

Dispersions of colloidal nanoparticles (NPs) in liquid crystals (LCs) attract a great deal of interest because of the potential for using them in reconfigurable self-assembled composite materials with unique physical behavior and properties [1-35]. From a fundamental standpoint of view, the elastic pair interaction forces between colloidal particles in LCs are typically measured by balancing them with the hydrodynamic Stokes drag [2-9], which makes it important to understand mechanisms governing the diffusivity of colloids in the host fluid related to the viscous drag through the Stokes-Einstein equation. On the other hand, Brownian motion of anisotropic colloids is of great fundamental interest by itself [1,36,37] and LCs provide the unique ability of exploring how such Brownian motion of anisotropic particles is affected by mechanical coupling between their orientation and director structures describing orientational ordering in these anisotropic fluids. Even spherical particles dispersed in LCs exhibit highly anisotropic diffusion properties that are dependent on the director field $\mathbf{n}(\mathbf{r})$ and topological defects induced by these particles, giving rise to effective anisotropy of nematic colloids, in addition to the strong anisotropy of viscous properties of LCs [2,38-40]. Translational diffusion of isometric dielectric nanospheres was studied in thermotropic [26,27] and lyotropic LCs [30]. It was found that the diffusivity of gold NPs can be affected by chemical functionalization of their surface [25] and the mobility of gold NPs on the LC-aqueous interface is significantly different relative to microparticles insensitive to the details of the interfacial environment [32]. The recent theoretical study [33] reports interesting results on the effect of anchoring type on diffusivity of NPs: NPs with homeotropic anchoring are expected to have lower diffusivity and diffusivity of NPs with planar anchoring does not change as compared to the case when neither homeotropic nor parallel anchoring is favored. The faceted shape of colloids offers additional director texturing mechanisms, as compared to those with smooth surface curvature [23]. Thus, these literature examples [8-21] show that the interplay of colloidal shape and medium anisotropies of NPs and the chemical functionalization of colloidal surfaces are expected to play key roles in defining their diffusion properties [1], but have not been systematically studied in the past.

In this article, using dark-field video microscopy, we probe translational and rotational diffusion of gold NPs of different shape and surface functionalization dispersed in the nematic LC bulk. We measure corresponding diffusion coefficients and show that their persistent strong anisotropy with respect to the far-field director $\mathbf{n}_0$ is dependent on the shape and surface anchoring of colloidal particles, which is because of determining equilibrium alignment of these anisotropic colloids in a medium with anisotropic properties and local particle-induced director distortions and defects. We analyze these findings by invoking the comparison to diffusion of anisotropic particles in isotropic fluids and shape-isotropic particles (spheres) in anisotropic LC fluids. We also estimate elastic torques on elongated NPs experiencing the rotational diffusion around the axis normal to $\mathbf{n}_0$. We show that functionalization of NPs with photosensitive azobenzene groups allows for *in situ* control of surface anchoring through *trans*-*cis* isomerization, which, in turn, changes their diffusivity, providing additional insights into the nature of dependence of particle diffusion properties on surface functionalization.

## II. EXPERIMENT

### A. Materials and sample preparation

We used a single-compound room temperature nematic LC 4-pentyl-4-cyanobiphenyl (5CB) from Frinton Laboratories, Inc., as a host for gold NPs of different anisotropic shapes, dimensions, and surface properties (Table I). Elongated convex pentagonal (CPNs) and concave starfruitlike (CSNs) nanoprisms, platelike nanoburst (NBN) nanoparticles, and small nanorods (GNR1) had NSol(alkyl acrylate) polymer capping [29,41] and were obtained from Nanopartz, Inc., in ethanol solution. The smallest gold nanorods with polystyrene (GNR2) and PEG-5CB (GNR3) capping [42,43] were provided in, respectively, dichloromethane and chloroform solution by Zubarev from Rice University. Gold nanorods with photoresponsive surface anchoring (GNR4) were synthesized using a seed-mediated method [34] and capped with azobenzene-containing ligands that allow for switching orientation of the easy axis at the LC-particle interface between tangential and homeotropic (details of nanoparticle synthesis can be found in Ref. [44], and the chemical surface modification as well as synthesis of the ligands will be reported elsewhere [45]).

The studied nanoparticles were dispersed in 5CB at low concentration (< 1000 ppm) to obtain well-separated colloidal NPs. Nanorods in dichloromethane and chloroform were added to LC directly but NPs initially in ethanol solution were first redispersed in toluene and then the toluene dispersion was added to LC. The capping of NPs was stable upon exchanging solvents and transferring into LC due to a strong thiol covalent bonding [43]. After the evaporation of solvents at elevated temperature, the resulting colloidal dispersions were sonicated in an ultra-sonic bath for 30-60 min at a temperature above the clearing point and subsequently quenched to the room temperature. Immediately, obtained dispersions of nanoparticles in LC were filled into the gap of thickness $d$ = 3 and 10 $\mu$m between two glass substrates set by glass microfibers. The actual $d$ was measured in empty cells using the interference method and spectrophotometer USB2000 (Ocean Optics). Confining substrates were dip coated in an aqueous (1 wt %) solution of [3-(trimethoxysilyl) propyl]octadecyl-dimethylammonium chloride (DMOAP) to induce homeotropic (normal to the surface) [46] alignment of the far-field director $\mathbf{n}_0$ of LC. We used unidirectionally rubbed thin films of spin-coated and baked polyimide PI2555 (HD MicroSystem) or aqueous (1 wt %) poly(vinyl alcohol) (Sigma-Aldrich) to achieve planar (tangential to the surface) alignment of LC and to define the in-plane orientation of $\mathbf{n}_0$. One of the two glass substrates in each experimental cell was 0.15-0.17 mm thick to enable observations with oil immersion objectives of high numerical aperture (NA) and magnification. The LC-colloidal dispersions were filled in by capillary action when in a liquid crystalline phase and sealed with epoxy glue. The prepared samples were stable against NPs aggregation and precipitation for at least 24–48 h, within which our experiments were carried out.

### B. Methods and techniques

Dark-field microscopy was used for optical observations and video tracking of NPs using a multimodal experimental setup [22,47] built around an Olympus inverted microscope X81. The use of an Olympus 100× oil objective with variable NA = 0.6-1.3 and a customized dark-field condenser U-DCW (NA = 1.2-1.4) enabled the high contrast in dark-field videomicroscopy of studied NPs. Brownian motion of these colloids at room temperature was recorded with a CCD camera (Flea, PointGrey) at a rate of 15 fps, and exact position of a single NP as a function of time was then determined from captured sequences of images using particle tracking plugins of ImageJ software. The accuracy of determination of the NPs' positions and orientations can somewhat deteriorate because of lower signal-to-noise ratio (because of scattering due to director fluctuations) and birefringence of the LC host medium. Partially polarized light scattered from NPs in the LCs undergoes double refraction and even small inhomogeneity of $\mathbf{n}(\mathbf{r})$ caused by thermal fluctuations [6] could potentially modify the propagation of its extraordinary mode and thus affect the precise determination of their position with an additional introduced error or shift proportional to the LC birefringence and thickness of the sample. To mitigate this problem in our experiments, we used unpolarized light illumination and detection in the dark-field video tracking, since the light propagation direction would be sensitive to **n**(**r**) change only for the extraordinary mode and the ordinary mode propagates along a fixed direction [6]. We determined the precision of our measurements using a technique similar to that described in Refs. [48,49], which involves video tracking the position of stationary NPs immobilized on the confining substrates of LC cells. Light scattered from NPs was passing through the bulk of the same LC cells as used in the experiments. The standard deviation of the measured positions was 2.5 nm, which allows concluding that the spatial position of NPs in our measurements can be determined with a precision of 3 nm or better and that the effects of birefringence and light scattering due to director fluctuations are inconsequential from the viewpoint of the conclusions of our work. The photoresponsive capping at the surface of GNR4 nanorods was activated by irradiation from an OmniCure S2000 spot UV system (Lumen Dynamics).

## III. RESULTS

### A. Diffusion of rodlike convex pentagonal nanoprisms aligned perpendicular to $\mathbf{n}_0$

The dark-field microscopy allows direct real-time visualization of colloidal particles smaller than the diffraction limit [21,26-28,30,32,45] due to the strong scattering of visible light from gold NPs [50] having the appearance of bright spots on a dark, uniform background. The size of these spots does not directly correspond to the actual dimensions of the NPs but represents their scattering cross section that, in addition to the particle size, is determined by diffraction-limited resolution [50]. Although the shape anisotropy of scattering spots becomes unrecognizable as the size of the NPs decreases significantly below the diffraction limit, the intensity of the scattering typically remains dependent on the polarization of incident light and often allows one to deduce the actual orientation of such NPs [29, 44].

CPNs have homeotropic anchoring on their surfaces (Table I) and align with their long axis *a* perpendicular to a substrate surface rubbing defined far-field director orientation $\mathbf{n}_0$ = {0, 0, 1} [Fig. 1(a)]. The symmetry of resulting director distortions **n**(**r**) around CPNs [Fig. 1(a)] is of “quadrupolar” type [3,6,27], with encircling half-integer disclination loop (often called “Saturn ring”) [26,29] of winding number -1/2. CPN drifts in the plane of the liquid crystal cell due to Brownian motion. The typical erratic trajectory obtained from 4300 frames of video tracking of CPN is shown in Fig. 1(b). At the same time, CPN can also freely rotate around its transverse (short) axis $c$ ($\|\mathbf{n}_0$) [Fig. 1(a) and insets in Figs. 1(c)-1(d)]. Using dark-field videomicroscopy tracking data, one can construct a histogram of displacements $\Delta = \mathbf{r}(t + \tau) - \mathbf{r}(t)$ (Fig. 1), translational or rotational, NP makes from the frame to frame over the elapsed time $\tau$ [8,25,27,40]. In experiments $\mathbf{r}$ was measured in $x$ ($\perp\mathbf{n}_0$), $y$ ($\perp\mathbf{n}_0$), $z$ ($\|\mathbf{n}_0$), $a$, or $b$ directions for translational diffusion and due to angle $\theta$ changes for rotational diffusion (note that the symmetry of the LC-NP composite with NPs following director orientation states in the LC host is uniaxial). As expected, experimentally obtained displacement distributions (Fig. 1) can be fit by a Gaussian function of the form [8,25,27,40]

$$P(\Delta|\tau) = P_0(\tau)\exp[-\Delta^2/(4D_\alpha\tau)], \tag{1}$$

where $P(\Delta|\tau)$ is the probability that over the time $\tau$ a nanoparticle will displace by $\Delta$, $P_0(\tau)$ is a normalization constant, and a value $4D_\alpha\tau$ determines the width of the distribution [40,51], where $D_\alpha$ is a diffusion coefficient and the subscript indices $\alpha = \|,\perp$ stand for translational diffusion along and perpendicular to $\mathbf{n}_0$, respectively, indices $\alpha = a,b$ stand for translational diffusion along longitudinal and transverse axes of the axially symmetric anisotropic NP, respectively, and $\alpha = \theta$ for rotational diffusion around $c$. The larger width of distribution defines a larger diffusion coefficient. Figures 1(c) and 1(d) show displacement distributions of the CPN nanoprism in the planar cell. The difference in width of distributions [Fig. 1(c)] corresponding to two orthogonal directions indicates that the diffusion of CPNs is anisotropic with respect to $\mathbf{n}_0$. Interestingly, this anisotropy also depends on the orientation of the CPN while it freely rotates around $\mathbf{n}_0$, with respect to the plane of the cell (in the plane orthogonal to $\mathbf{n}_0$). When $a$ is roughly parallel to the plane of the cell [Fig. 1(c)], the diffusion in the direction normal to $\mathbf{n}_0$ is easier than along $\mathbf{n}_0$ ($D_\|/D_\perp < 1$), but diffusivity anisotropy switches to $D_\|/D_\perp > 1$ when $a$ is normal to the plane of the cell [Fig. 1(d)]. Because of the long-range orientational ordering of the nematic LC host, this correlation persists over long periods of time and is rather unique for anisotropic nematic fluids. However, the limitations of dark-field microscopy observations of the particle motion corresponding to all five degrees of freedom preclude accurate characterization of CPN's rotational diffusion in the plane orthogonal to $\mathbf{n}_0$ in this cell geometry, which in principle could be quantitatively correlated to the translational diffusion.

To probe rotational diffusion of CPNs, we observed them in a homeotropic cell [inset of Fig. 1(e)]. The angular histogram [Fig. 1(e)] shows orientations of *a* acquired for ~ 5 min. The rotational diffusion coefficient of CPN is found from the distribution of angular displacements $\Delta\theta = \theta(t+\tau) - \theta(t)$ [Fig. 1(e)] to be equal $D_\theta = 0.028\ \mathrm{s}^{-1}$. The translational diffusion of CPN in the homeotropic cell is isotropic ($D_y/D_x \approx 1$) in the laboratory frame [Fig. 1(f)]. However, the role of the particle's shape anisotropy becomes apparent once we convert measured displacements $\Delta_n(x, y)$ to displacements $\Delta_n(a, b)$ relative to the rotating body frame of the nanoparticle with axes *a* and *b* [Fig. 1(a)] using the common coordinate transformation [36,37]

$$\begin{pmatrix} \Delta_{an} \\ \Delta_{bn} \end{pmatrix} = \begin{pmatrix} \cos\theta_n & \sin\theta_n \\ -\sin\theta_n & \cos\theta_n \end{pmatrix} \begin{pmatrix} \Delta_{xn} \\ \Delta_{yn} \end{pmatrix}, \tag{2}$$

where $\theta_n = [\theta(t_{n-1}) + \theta(t_n)]/2$. Clearly, the diffusivity of CPN appears anisotropic ($D_a/D_b > 1$) relative to the body frame [Fig. 1(g)]. Although it would be certainly of interest, we could not characterize orientational fluctuations of CPNs out of the plane perpendicular to $\mathbf{n}_0$ because the available resolution was insufficient to decouple the rotational diffusions of CPNs around their transverse axes parallel and perpendicular to $\mathbf{n}_0$ in a planar cell.

### B. Diffusion of rodlike concave starfruit-shaped nanoprisms aligned parallel to $\mathbf{n}_0$

CSN nanoprisms with concave faceted faces align along $\mathbf{n}_0$ [Fig. 2(a) and 2(b)]. Director distortions $\mathbf{n}(\mathbf{r})$ have "quadrupolar" symmetry with two surface point defects called "boojums" at both ends [26,29]. In dark-field microscopy observations, CSNs appear as bright elongated spots that, on average, align along $\mathbf{n}_0$ to minimize the free energy due to particle-induced elastic distortions and surface energy of anisotropic molecular interactions at particle surfaces. The anisotropy of translational diffusion of CSNs is large ($D_\parallel/D_\perp > 2$) (Table I) as they move much more rapidly along $\mathbf{n}_0$ than in other directions [Fig. 2(b)]. CSNs also show distinct rotational fluctuations around the transverse short axis ($b\perp\mathbf{n}_0$). Using standard tracking plugins of ImageJ, one can measure the orientation of CSN in each frame with accuracy $\approx 0.01$ rad (0.5°). Figure 2(c) shows that the orientation angle $\phi$ of CSN fluctuates around $\phi_0 = 0$, which corresponds to the orientation along $\mathbf{n}_0$. Since the size of the particle is comparable to $K/W_a$, where $K$ is an average Frank elastic constant and $W_a$ is a surface anchoring coefficient [6], in general, one can expect that the deviation of the particle's long axis away from the equilibrium orientation at $\phi_0 = 0$ is associated with both elastic and surface anchoring energy costs. However, because a simple analytical treatment of this finite-anchoring complex-geometry problem is complicated, it is instructive to assume that the nanoprism has a cylindrical shape, with the cylinder circumscribing the nanoprism, and consider the regimes of strong and weak anchoring separately. The free energy cost due to particle rotation away from the ground-state orientation is of purely elastic origin in the strong anchoring case and can be described by an elastic potential $U_e(\phi) = \kappa_e\phi^2$ [11,52] due to the additional director distortions caused by the elongated NP with planar surface anchoring deviating from the orientation $\phi_0$ set by $\mathbf{n}_0$, where $\kappa_e$ is the stiffness coefficient due to the elastic energy cost of angular rotations. In the weak anchoring regime, on the other hand, one can assume that there are no particle-induced elastic distortions around the colloids before and after the rotation but the director field meets the particle surface at different angles upon being rotated, thus resulting in an additional increased surface anchoring energy cost associated with this rotation. Assuming Rapini-Papoular anchoring potential [6], one can roughly estimate this surface anchoring energy cost as $\pi RLW_a\sin^2\phi$, which in the limit of small angles can be expressed as $U_a(\phi) \approx \kappa_a\phi^2$, where $\kappa_a$ is the orientational trap's stiffness coefficient due to the anchoring energy cost. In the intermediate regime of finite anchoring, one deals with a combination of elastic distortions and deviations of the director from its easy axis at the nanoparticle's surface and thus one can introduce $\kappa_{ea}$, the orientational trap's stiffness coefficient due to both the anchoring and elastic energy costs. The probability distribution of angles $\phi$ between the axial direction of a concave nanoprism and $\mathbf{n}_0$ [Fig. 2(d)] can be described by a function $f(\phi) \sim \exp[-\kappa\phi^2/k_BT]$, where $k_B$ is the Boltzmann constant and $T = 300$ K [6,51]. Fitting the experimental distribution [Fig. 2(d)] with $f(\phi)$, we find the standard deviation of CSN's orientation from $\mathbf{n}_0$ as $\langle\phi\rangle \approx \pm(0.035 - 0.042)$ rad [$\approx \pm(2 - 2.5)°$] at room temperature and experimental trap stiffness $\kappa = (1.67 \pm 0.05)\times10^{-18}$ N m that could correspond to the different above-mentioned anchoring regimes.

Using orientation fluctuation data [Fig. 2(c)], we calculate the mean square angular displacement $\langle\Delta\phi^2\rangle$ as a function of elapsed time $\tau$ [Fig. 2(e)]; note that there is a constant shift of $\langle\Delta\phi^2\rangle$ at $\tau = 0$ s resulting partly from the image pixels noise. At short times $10^{-3}$ s $< \tau < 10^{0}$ s $\langle\Delta\phi^2\rangle$ increases almost linearly and saturates at $\sim 0.00315$ rad$^2$ after $\tau \sim 7$ s. The CSN's orientational fluctuations can be described by a Langevin equation for rotational motion,

$$I\phi''(t) = T_{\phi}(t) - \zeta\phi'(t) - \partial_{\phi}U(\phi), \tag{3}$$

where $I$ is the moment of inertia, $\zeta$ is a viscous rotational friction coefficient, $\partial_{\phi}U(\phi) = 2\kappa\phi$ is a restoring torque [11,52] on the elongated nanoprism not aligned along $\mathbf{n}_0$, and $\kappa$ is a rotational stiffness of the orientational trap that constrains rotations of the particle away from $\mathbf{n}_0$. A rapidly fluctuating torque $T_{\phi}(t)$ is due to unceasing random collisions of the LC molecules with the nanoparticle and $\langle T_{\phi}(t)\rangle = 0$, $\langle T_{\phi}(t)T_{\phi}(t')\rangle = 2\zeta k_B T\delta(t - t')$ [53, 54]. The Brownian motion in our system is overdamped, as the Reynolds number is very small ($Re \sim 10^{-8} << 1$) [29], so that the inertia term can be neglected and the solution for $\langle\Delta\phi^2\rangle$ [53, 54] can be found as

$$\langle\Delta\phi^2(t)\rangle = (k_B T/2\kappa)(1 - \exp[-4\kappa t/\zeta]). \tag{4}$$

By fitting the measured $\langle\Delta\phi^2\rangle$ data [Fig. 2(e)] with Eq. (4), one can extract the viscous rotational friction coefficient $\zeta = (18.37 \pm 0.43)\times10^{-18}$ N m s and the stiffness coefficient $\kappa = (1.73 \pm 0.01)\times10^{-18}$ N m. At short times $\tau << \zeta/4\kappa$ ($\zeta/4\kappa \approx 2.6$ s), free diffusion dominates [Fig. 2(e)], $\langle\Delta\phi^2(t)\rangle = 2(k_B T/\zeta)t$, and the rotational diffusion coefficient around the transverse axis is $D_{\phi} = k_B T/\zeta = 2.3\times10^{-4}$ s$^{-1}$ for the CSN nanoprism. Nematic elasticity and surface anchoring dominate at long times ($\tau >> 2.6$ s) and thermal energy is too small ($k_B T/2\kappa \approx 10^{-3}$) to rotate elongated CSN nanoprism out of the elastic orientational trap set along $\mathbf{n}_0$. Values of $\kappa$ extracted from distribution [Fig. 2(d)] and mean squared angular displacement [Fig. 2(e)] are in a good agreement with each other. Using the expression for the stiffness coefficient $\kappa = 2\pi CK$ for a cylindrical particle in nematic [11,52], where $K \approx 6\times10^{-12}$ N for 5CB [55], $C = 2L\beta/\ln[(1 + \beta)/(1 - \beta)] \approx 1.7\times10^{-7}$ m is a capacitance of cylindrical particle, $\beta = [1 - (R/L)^2]^{-1/2}$ and $L$ is a length and $R$ is a radius [52], one can roughly estimate it as $\kappa \approx 5.22\times10^{-18}$ N m, which is of the same order of magnitude but several times larger than values extracted from experimental data [Fig. 2(e)]. This difference between estimated and extracted values of $\kappa$ can be attributed to the difference between $C$ of assumed cylindrical shape and experimental complex shape of a concave nanoprism CSN, neglecting the effects of confinement and capping, and to experimental uncertainty in measured $L$ and $R$ of nanoprisms. On the other hand, using experimental $\kappa = (1.67 \pm 0.05)\times10^{-18}$ N m, it is possible to determine an apparent capacitance for a measured CSN as $C_{\text{app}} \approx (0.53 \pm 0.02)\times10^{-7}$ m, which is of the same order of magnitude but somewhat smaller than the theoretically estimated $C$. Using experimentally obtained $\kappa$, one can also estimate elastic energy of additional distortions caused by deviation of CSN from $\mathbf{n}_0$ on average by $\langle\phi\rangle$ [Fig. 2(d)] as $U(\phi) = \kappa\langle\phi\rangle^2 \approx (2.1\text{-}3)\times10^{-21}$ J, which is comparable to thermal energy $k_BT$. The corresponding elastic restoring torque on a CSN nanoparticle is $\partial_\phi U(\phi) = 2\kappa\langle\phi\rangle \approx (1.2 - 1.5)\times10^{-19}$ N m.

Since the alignment of concave nanoprisms in a nematic sample is determined by finite nematic elasticity and/or surface anchoring, the order parameter of CSN aligned in a nematic sample and deviating from $\mathbf{n_0}$ by $\phi$ can be calculated as $S = \int_0^\pi P_2(\cos\phi)\, f_e(\phi)\sin\phi\, d\phi \approx 0.996$ [6, 51], where $P_2(\cos\phi)$ is a second Legendre polynomial and $f_e(\phi) \sim \exp[-\kappa_e\phi^2/k_BT]$ is calculated with experimentally determined $\kappa_e$ [Fig. 2(d)]. It is comparable to the order parameter $S \approx 0.998$ calculated for a cylindrical particle with similar dimensions (Table I) and $f_e(\phi) \sim \exp[-2\pi CK\phi^2/k_BT]$. On the other hand, assuming a weak anchoring regime, substitution of $f_s(\phi) \sim \exp[-\pi RLW_a\sin^2\phi/k_BT]$ into the equation for the order parameter with experimentally determined value $S = \int_0^\pi P_2(\cos\phi)\, f_s(\phi)\sin\phi\, d\phi \approx 0.996$ allows for estimation of the anchoring coefficient at the surface of CSN as $W_a = 2.2\times10^{-5}$ J/m$^2$, which is comparable to typical values measured for thermotropic LC-solid interfaces [6], and an anchoring extrapolation length as $K/W_a \approx 230$ nm. We note that these calculations assuming both regimes give only an order of magnitude estimates because we in fact have finite surface anchoring conditions. However, the weak anchoring conditions in similar situations can be assumed when the particle size is further decreased as compared to $K/W_a$ and the strong anchoring regime can be assumed for larger cylindrical particles, so that similar measurements done for anisotropic colloids of different size in LCs may allow for a quantitative characterization of LC elastic and anchoring properties.

### C. Diffusion of disklike nanoparticles aligned perpendicular to $\mathbf{n}_0$

NBN nanoplatelets are shaped as disks with irregular edges and align with their flat large-area faces normal to $\mathbf{n}_0$ [Fig. 3(a)], which is due to homeotropic surface anchoring (Table I) [29]. Similar to CPNs, director distortions around NBNs are quadrupolar with a disclination loop encircling their edges. As a result, NBNs are oriented parallel to substrates in homeotropic cells [inset of Fig. 3(b)] and edge-on in planar cells [inset of Fig. 3(c)]. NBNs are free to rotate about $\mathbf{n}_0$ as well as have their normal to large-area faces orientationally deviate from $\mathbf{n}_0$ under the influence of thermal fluctuations. However, the small shape anisotropy of NBNs does not give appreciable difference in dark-field images corresponding to different orientations accessed by particles and we, therefore, could not characterize this rotational diffusion in our experiments. Using data on translational diffusion of NBNs in both types of cells, we calculate their mean square displacements (MSDs) $\langle \Delta x^2 \rangle$, $\langle \Delta y^2 \rangle$ and $\langle \Delta z^2 \rangle$ in three mutually orthogonal directions, $x$, $y$ ($\perp\mathbf{n}_0$), and $z$ ($\|\mathbf{n}_0$), as a function of $\tau$, finding that the MSD increases linearly for all three directions [Figs. 3(b) and 3(c)]. Consistent with the geometry of our experiments, diffusion of NBNs is isotropic in a homeotropic cell [Fig. 3(b)] but strongly anisotropic in a planar cell [Fig. 3(c)]. We fit experimental data with a linear equation for free diffusion [51,56] $\langle \Delta \mathbf{r}^2 \rangle = 2D_\alpha \tau$ and find the diffusion coefficients for NBN nanoplatelets $D_\perp = D_x \approx D_y$ and $D_\| = D_z$, $D_\| > D_\perp$ (Table I).

### D. Diffusion of rodlike nanoparticles with varying surface boundary conditions and equilibrium alignment along $\mathbf{n}_0$

We also studied diffusion of small ($L < 70$ nm and $2R < 30$ nm) gold nanorods (GNRs) with different surface chemistry and anchoring conditions (Table I). Their alignment with respect to $\mathbf{n}_0$ and director distortions that they induce were studied in detail in Ref. [29] using polarization dependent two-photon luminescence (TPL) [57]. These past studies showed that GNR2 nanoparticles with polystyrene capping exhibit planar surface anchoring and align parallel to $\mathbf{n}_0$, each creating director distortions in the form of an elastic quadrupole and with two boojums at their ends [Fig. 4(a)] [29]. Chemical capping of nanorods GNR1 and GNR3 causes homeotropic anchoring of LC molecules at their surface and results in colloidal elastic quadrupoles with half-integer disclination loops surrounding them [Fig. 4(c)]. In dark-field microscopy, these nanorods appear as tiny bright spots undergoing diffusion [inset of Fig. 4(b)].

Since the resolution of dark-field video tracking is not sufficient to probe orientational fluctuations of these small nanorods, we will focus exclusively on their translational diffusion. Figures 4(b) and 4(d) show data obtained in planar nematic cells for two orthogonal directions, indicating that the MSD is linearly increasing with elapsed time for GNRs with tangential and homeotropic alignment. The mobility of GNRs is anisotropic with respect to $\mathbf{n}_0$. The ratio of self-diffusion coefficients deduced from fitting the MSD data is within $1 < D_{\|}/D_{\perp} < 2$ (Table I). The measured diffusion parameters are of the same order of magnitude as those recently reported for spherical gold [25] and silica nanoparticles [27] and calculated for prolate spheroids for an isotropic fluid of comparable viscosity (Table I). A comparison of the diffusion data for studied nanorods indicates the important role that the surface anchoring plays in determining diffusion properties, which is because of its role in defining the alignment of rodlike nanoparticles with respect to the far-field director [6,11] and because of controlling the "corona" of induced elastic distortions and defects around them.

To further explore the role of surface anchoring boundary conditions in determining anisotropic NP diffusion properties, we also used small nanorods GNR4 with capping incorporating photosensitive azobenzene groups that allow for optical control of surface anchoring. These azobenzene-containing ligands undergo *trans*-*cis* isomerization under the UV light illumination as shown in Fig. 5(a). Without UV light, the azobenzene units are in the *trans*-resting state promoting the homeotropic surface anchoring of LC molecules [Fig. 5(b)] [58]. However, when the sample is illuminated with the UV light, the azobenzene moieties transition into the *cis*-state [Fig. 5(a)] promoting tangential alignment of LC molecules [Fig. 5(b)] so that the orientation of the surface groups is matching the surrounding LC alignment. The transformation is reversible; the azobenzene units go back to the *trans*-state under the ambient light (and microscope illumination) shortly after the UV is off. This photoinduced change in surface anchoring properties of GNR4s alters their Brownian motion. Figure 5(c) shows measured MSD data for nanorods before illumination. Diffusion is anisotropic with respect to $\mathbf{n}_0$ ($1 < D_{\|}/D_{\perp} < 2$) with parameters comparable to other GNRs with homeotropic anchoring and similar alignment (Table I). However, under the unpolarized UV illumination the diffusion along $\mathbf{n}_0$ significantly increases [Fig. 5(d)] resulting in the increased anisotropy of mobility ($2 < D_{\|}/D_{\perp} < 3$). After UV is off, the azobenzene units go back to the *trans*-state and GNR4 show diffusivity similar to the one before irradiation [Fig. 5(c)]. The increased diffusivity cannot be attributed just to the increased temperature of NPs and neighborhood as a result of their UV absorption because their diffusivity increases significantly mostly in one direction along $\mathbf{n}_0$. Furthermore, we did not observe such dramatic effect on diffusion when using NPs without azobenzene-containing ligands. Our MSD data [Figs. 5(c) and (d)] are in good qualitative agreement with the theoretical study of Moreno-Razo *et al.* [33] showing that diffusivity of NPs with homeotropic anchoring is lower than that of NPs with planar anchoring. The value of translational diffusion coefficients is inversely proportional to the NP diameter [56], which can indicate that NPs have larger "apparent" dimensions due to defects and strong elastic distortions expected for homeotropic anchoring boundary conditions [33].

## IV. DISCUSSION

Experimentally obtained results for anisometric nanocolloids (Table I) show that their self-diffusion and dynamic behavior in nematic LCs strongly depends on size, shape, and surface functionalization. While translational self-diffusion of NPs in LCs, on average, always increases with decreasing their size, as expected [51], the anisotropy of this diffusion ($D_{\|}/D_{\perp}$) with respect to the far-field director changes not only due to their shape but also surface anchoring and the ensuing elasticity- and surface-anchoring-mediated alignment (Table I). The combination of anisotropies of particle shape, host medium, and LC-colloidal interfacial interactions yields diffusion properties that are fairly unique to nematic nanocolloidal soft matter systems. For example, elongated NPs aligned parallel to $\mathbf{n}_0$ tend to diffuse more rapidly along $\mathbf{n}_0$ than in directions perpendicular to it, just as one would expect for diffusion of elongated objects along their long body axis, due to lower resistance of the surrounding medium as well as due to the anisotropic viscosity of the LC. However, the comparison of diffusion anisotropies exhibited by different elongated NPs (Table I) shows that this anisotropy is also influenced by anisotropy of the "corona" of elastic distortions and defects induced by these colloids, which are dependent on surface anchoring and particle dimensions. Furthermore, diffusion of elongated particles aligned perpendicular to the far-field director ($a\perp\mathbf{n}_0$), such as CPNs, is coupled to their orientational thermal fluctuations and, at certain orientations [Fig. 1(c)], is faster in the direction perpendicular to $\mathbf{n}_0$ but the opposite is true for other orientations (Table I). Similar rich diffusion properties are also exhibited by oblate particles (Table I). To elucidate the origins of this rich physical behavior, we compare our experimental results for anisotropic NPs in anisotropic hosts to theoretical and experimental results for anisotropic NPs in isotropic hosts and isotropic spherical colloids in anisotropic host fluids (Table I).

It is instructive to first compare diffusion of anisotropic NPs in an anisotropic nematic LC with that in a conventional isotropic fluid. Assuming that the used NPs can be approximated as spheroids of length $L$ and radius $R$, their translational diffusion coefficients in isotropic fluid can be calculated with respect to their rotational symmetry axis using known equations [59] for geometrical parts of the corresponding drag coefficients,

$$\begin{aligned} D_a &= \frac{k_B T}{8\pi R\alpha_4\left(p^2-1\right)}\left\{\frac{\left(2p^2-1\right)\ln\left[p+\sqrt{p^2-1}\right]}{\sqrt{p^2-1}}-p\right\} \\ D_b &= \frac{k_B T}{16\pi R\alpha_4\left(p^2-1\right)}\left\{\frac{\left(2p^2-3\right)\ln\left[p+\sqrt{p^2-1}\right]}{\sqrt{p^2-1}}+p\right\} \end{aligned}, \quad \text{for } p>1 \qquad (5)$$

and

$$\begin{aligned} D_a &= \frac{k_B T}{8\pi R\alpha_4\left(1-p^2\right)}\left\{\frac{\left(1-2p^2\right)\arccos p}{\sqrt{1-p^2}}+p\right\} \\ D_b &= \frac{k_B T}{16\pi R\alpha_4\left(1-p^2\right)}\left\{\frac{\left(3-2p^2\right)\arccos p}{\sqrt{1-p^2}}-p\right\} \end{aligned}, \quad \text{for } p<1 \qquad (6)$$

where $p = L/2R$. Since the studied anisotropic NPs tend to align with respect to the far-field director so that their long and short axes are either perpendicular or parallel to $\mathbf{n}_0$, these formulas can be used for comparison with experimental diffusion coefficients in directions parallel or perpendicular to $\mathbf{n}_0$. For rough estimates, while neglecting the anisotropy of viscous properties of the fluid host, the effective viscosity can be approximated by the Leslie coefficient $\alpha_4$ = 0.075 Pa s of the studied 5CB [4,6,8]. While the calculated and measured values (Table I) have the same order of magnitude, one can also see that the nematic anisotropy in terms of both viscous properties and surface anchoring properties changes and enriches the diffusivity of the CPNs as compared to isotropic fluids. CPNs in a homeotropic cell [Figs. 1(e) and 1(g)] exhibit directly probed translational diffusion that is easier in the direction of their long axis, which is qualitatively consistent with calculations for an isotropic host (Table 1). On the other hand, the study of the diffusivity of CPN freely rotating around $\mathbf{n}_0$ in a planar cell can be modeled as diffusion of an effective prolate spheroid that one obtains by rotating the CPN around $a$. The estimates of diffusion coefficient for such a particle are again consistent with the observation that CPN is diffusing more easily in a direction perpendicular to $\mathbf{n}_0$ than parallel to $\mathbf{n}_0$. More interestingly, although the motion of an infinitely thin disk or oblate spheroid in isotropic fluids (Table I) is expected to be easier edgewise [56,59], elastic quadrupoles created by NBNs and surrounding director distortions and defects in anisotropic nematic hosts move more rapidly along $\mathbf{n}_0$, which is normal to the flat plane of nanoplatelets (Table I). This difference again demonstrates the important role that is played by nematohydrodynamics, elastic distortions, and disclination defects induced by the particle, and elastic coupling [30,38] of the NP axes to the rotational symmetry of nematic LCs. An additional factor that could be important to account for is the strongly irregular edge shape of NBNs [29] and small aspect ratio $L/2R \approx 0.2$ for which the used theoretical results are expected to give only very rough estimates (Table I). These findings are natural as even spherical particles dispersed in LCs were previously found to diffuse anisotropically due to the host medium's anisotropy and formation of defects and coronas of director distortions.

Some of the most interesting observations that arise due to the interplay of particle and host medium anisotropies include the following. Strong particle shape coupling to the symmetry and ordering of the nematic host [30] leads to increased diffusion anisotropy ($D_{\parallel}/D_{\perp} > 2$) for concave CSN nanoprisms, NBN nanoplates, and UV irradiated nanorods GNR4$_{cis}$, being higher than that previously characterized for both isometric spherical colloids in nematic hosts [26,27,39,40,60-62] and prolate [Eq. (5)] and oblate [Eq. (6)] spheroids with similar aspect ratios in isotropic fluids (Table I). Importantly, the coupling to the nematic symmetry can reverse the diffusion anisotropy from $D_a/D_b < 1$ to $D_a/D_b > 2$ as compared to what is expected for particles with similar shapes dispersed in isotropic hosts (see the example of NBNs in LC and oblate spheroids in isotropic fluids compared in the Table I). The rotational diffusion is also anisotropic and depends on the orientation of the rotation axis with respect to $\mathbf{n}_0$ [compare Figs. 1(e) and 2(c)-2(e)]; it is faster around $\mathbf{n}_0$ [Figs. 1(a) and 1(e)]. The rotational diffusion around the axis normal to $\mathbf{n}_0$ is strongly hindered or bound by the LC elasticity [Figs. 2(a) and 2(e)]. Often, as in the example of CPNs, the anisotropy of translational diffusion is coupled to the rotational diffusion [Figs. 1(c) and 1(d)]. Furthermore, translational self-diffusion of nanorods can be significantly altered *in situ* using the capping with photosensitive and mesogenic ligands (Fig. 5).

Even though the exploration of many interesting features of diffusion of anisotropic nanocolloids in the nematic LC in our experiments was limited by resolution of optical microscopy and accessibility of all the translational and rotational degrees of freedom for characterization, obtained results already show many unexpected and fundamentally interesting properties. Theoretical analysis of the effective viscous drag coefficient for nematic colloids with director distortions around them was previously studied only for spherical particles, and our study shows that there is a need of exploring further, both theoretically and experimentally, the role that the geometric shape of particles plays in defining diffusion properties of nematic nanocolloids.

## IV. CONCLUSIONS

We characterized the translational and rotational diffusion of gold NPs dispersed in a nematic LC, while varying their size, shape, and surface anchoring condition. Experimental data show strong anisotropy of translational diffusion with respect to the director $\mathbf{n}_0$, which is dependent on the alignment of anisotropic colloids as well as their shape and surface anchoring. Elastic torques on aligned NPs influence their rotational diffusion, making it very different from that observed in isotropic fluids. Furthermore, functionalization of NPs with photosensitive azobenzene groups allows for *in situ* control of NPs' surface anchoring through *trans-cis* isomerization, which in turn changes their diffusivity. Our experimental findings pose challenges for theoretical modeling of nanocolloidal diffusion and self-assembly in LCs; additionally they may impinge on applications of nanoparticle-LC composites in nanophotonics and nanoscale energy conversion. Although we focused on uniaxially symmetric nematic fluid host and colloidal particles with similar symmetry, it will be of interest to extend these studies to both phases and NPs of lower symmetry. In particular, it will be of interest to extend our work to other mesomorphic phases, such as cholesteric, smectic, and columnar LCs, which will allow one to probe the role of chirality and intrinsic quasi-long-range or long-range partial positional ordering in determining diffusivity of colloidal NPs of various shapes and sizes.

## ACKNOWLEDGMENTS

This work was supported by the International Institute for Complex Adaptive Matter (B.S.) and the NSF Grant No. DMR-0847782 (D.G.) and DOE Grant No. ER46921 (B.S. and I.I.S.). We thank Christian Schoen and Shelley Coldiron from Nanopartz, Inc., Eugene R.

Zubarev, Pramit Manna, Leonid Vigderman from Rice University, and Qingkun Liu from University of Colorado for providing the studied nanoparticles, as well as Jianqiang Zhang from Fudan University and University of Colorado at Boulder for help with surface functionalization of some of the nanoparticles. We also acknowledge discussions with Paul Ackerman, Julian Evans, Clayton Lapointe, Taewoo Lee, Qingkun Liu, Angel Martinez, Holger Stark, Rahul Trivedi, and Jianqiang Zhang.

# Figures

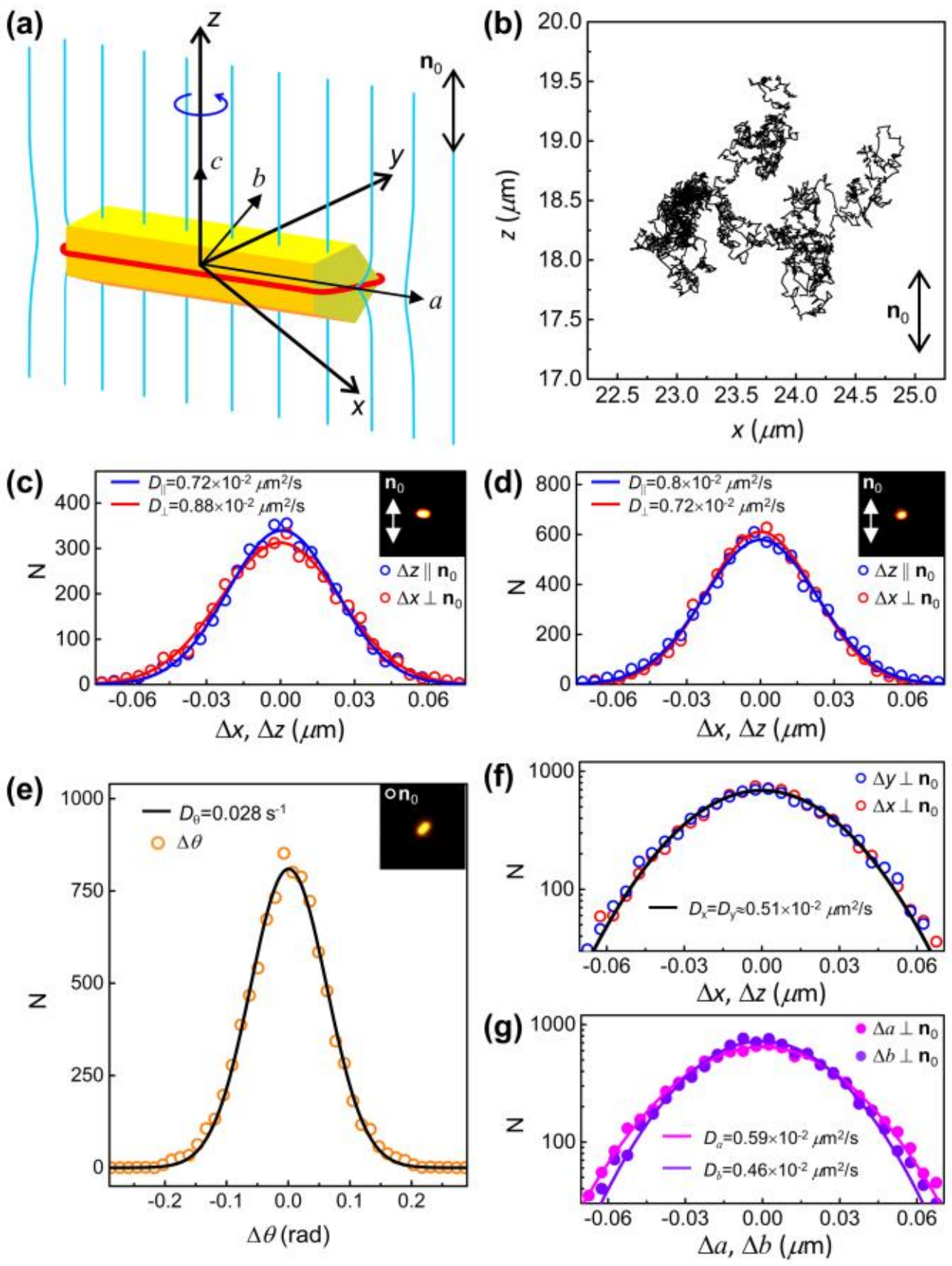

FIG. 1. (Color online) Diffusion of CPN nanoparticle in a nematic liquid crystal. (a) Schematic diagram of CPN and surrounding director field **n(r)** (thin blue lines); a thick red line around NP shows a disclination loop. (b) Trajectory of Brownian motion in the planar cell ($d \approx 3$ $\mu$m). (c), (d) Histograms of displacements along (*z* axis) and perpendicular (*x* axis) to the far-field director $\mathbf{n}_0$ for CPN oriented in the plane (*x* axis) (c) and out of the plane (between *z* and *x* axes) (d) of a planar cell; insets show dark-field textures of corresponding CPNs. The size of insets is $8 \times 8$ $\mu$m$^2$. Solid lines are a fit with Eq.(1). (e) Histograms of angular displacements of CPN around $\mathbf{n}_0$ collected for ~ 10 min in a homeotropic cell ($d \approx 10$ $\mu$m). Black solid line is a fit with Eq.(1). (f), (g) Histograms of displacements along *x* and *y* axes (f) and *a* and *b* axes of a body frame (g) in a homeotropic cell. Displacement data were collected at a rate of 15 fps.

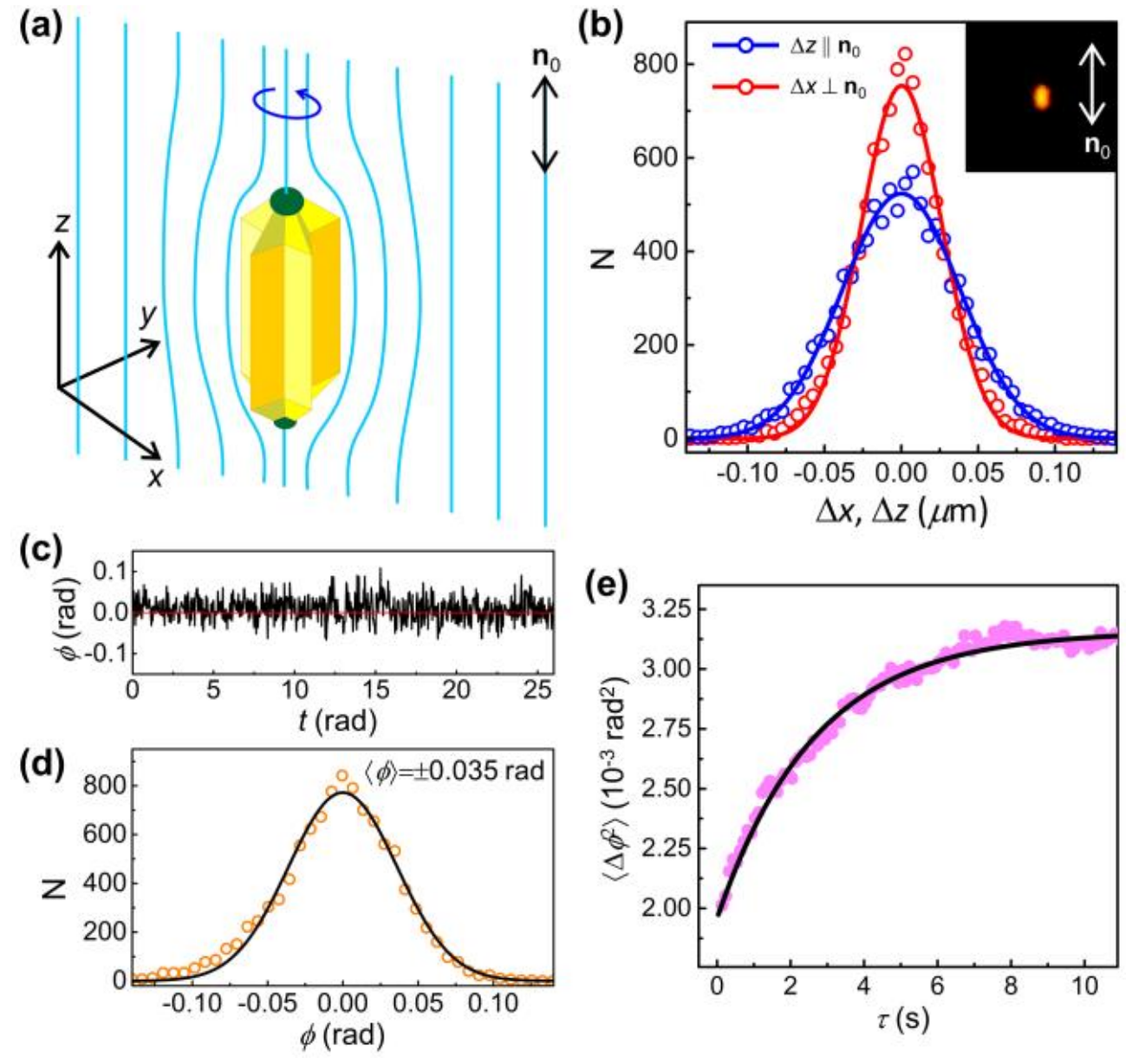

FIG. 2. (Color online) Diffusion of CSN nanoparticle in a nematic liquid crystal. (a) Schematic diagram of CSN and surrounding director field **n(r)** (thin blue lines). Green filled spheres at the ends of NP represent surface point defects boojums. (b) Histograms of displacements along ($z$ axis) and perpendicular ($x$ axis) to $\mathbf{n}_0$ in a planar cell ($d \approx 10$ $\mu$m). Solid lines show a fit to the data (open symbols) using Eq.(1). Inset shows a corresponding dark-field $8 \times 8$ $\mu\text{m}^2$ texture of CSN. (c) Experimental data showing the orientation $\phi$ of elongated CSN nanoprism with respect to $\mathbf{n}_0$ (a thin red line at $\phi_0 = 0°$) as a function of time at room temperature. (d) Histogram of angles $\phi$ between the orientation of CSN long axis and $\mathbf{n}_0$ collected at 15 fps. A solid line is a Gaussian fit. (e) Mean square angular displacements $\langle \Delta\phi^2 \rangle$ as a function of $\tau$ in a planar cell. A solid line is a fit to the data (filled circles) using Eq.(4).

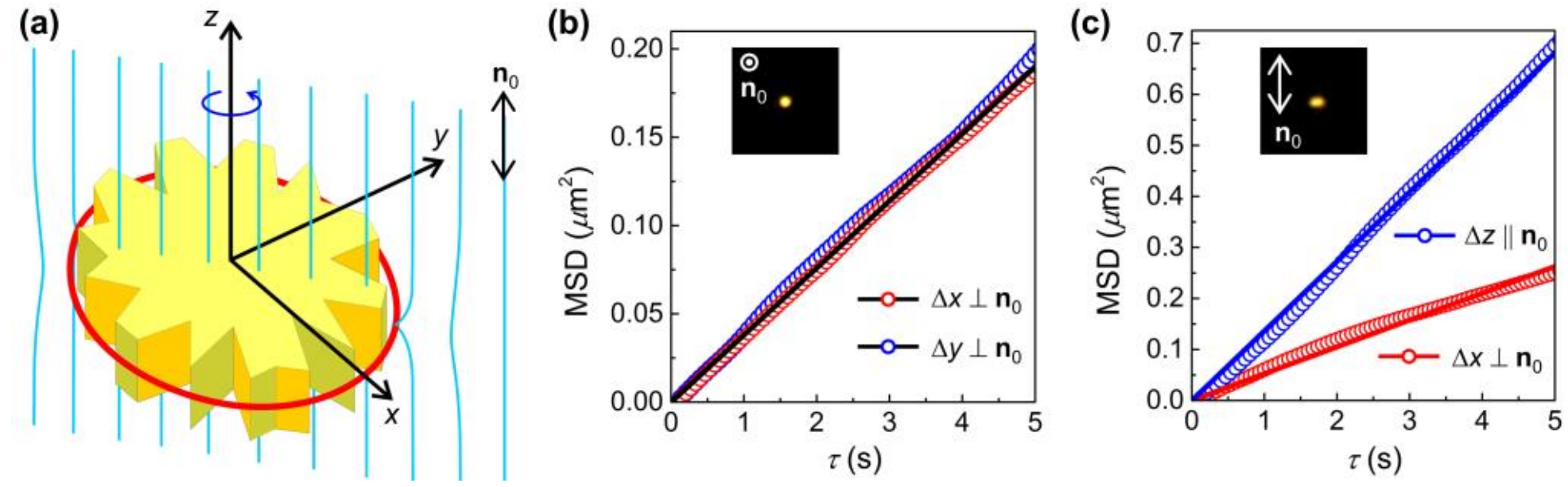

FIG. 3. (Color online) Diffusion of NBN nanoparticle in a nematic liquid crystal. (a) Schematic diagram of NBN and surrounding director field **n(r)** (thin blue lines). (b) Mean square displacements along $x$- and $y$-axes in a homeotropic cell ($d \approx 3$ $\mu$m) as a function of $\tau$. (c) Mean square displacements $\langle \Delta z^2 \rangle$ and $\langle \Delta x^2 \rangle$, respectively, parallel and perpendicular to $\mathbf{n}_0$ in a planar cell ($d \approx 10$ $\mu$m). Solid lines are a linear fit to the data (open symbols). Insets show dark-field textures of NBNs in corresponding geometries. The size of insets is $8 \times 8$ $\mu\text{m}^2$.

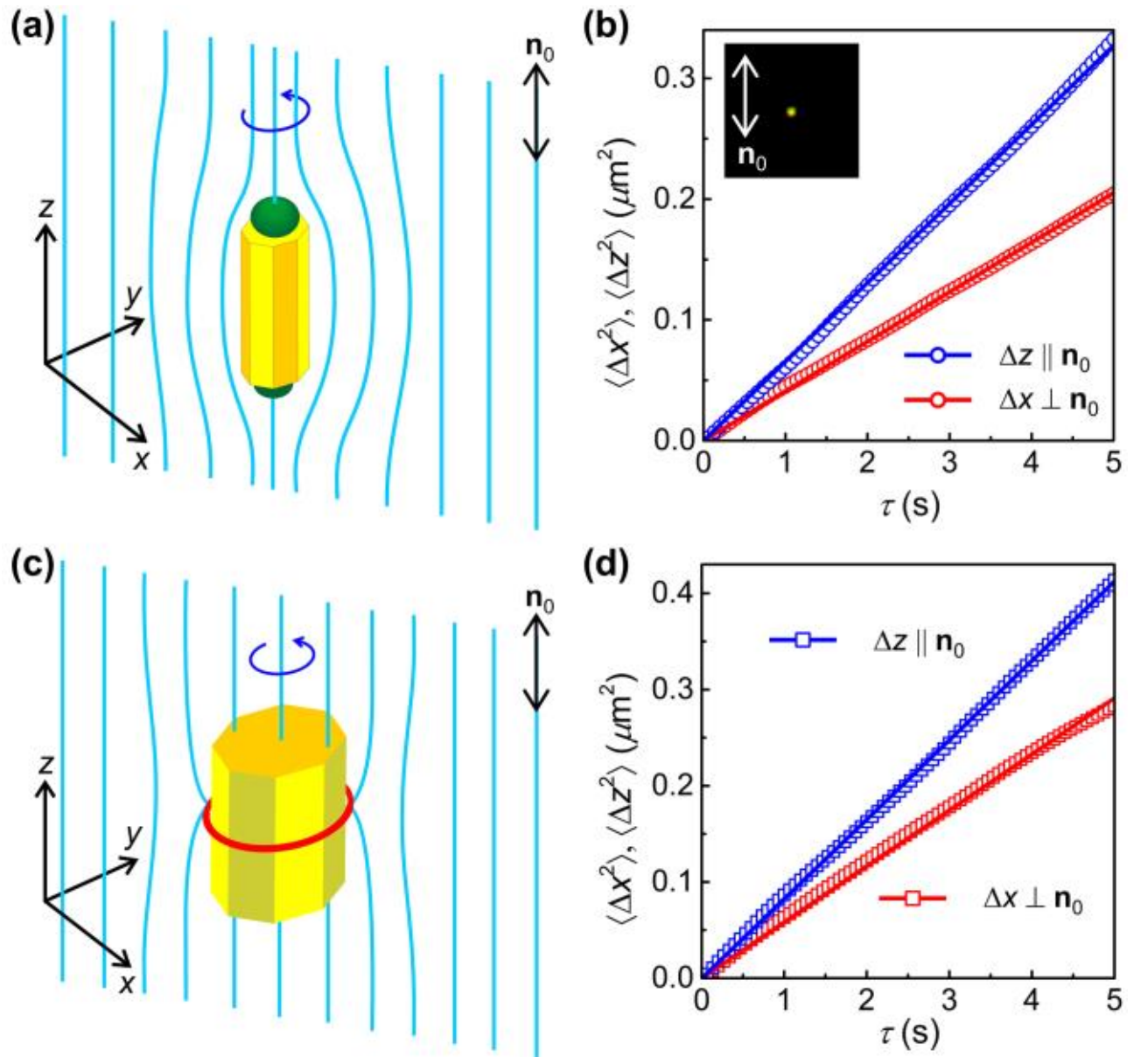

FIG. 4. (Color online) Diffusion of NPs with planar and homeotropic surface anchoring in a nematic liquid crystal. (a) Schematic diagram of GNR2 nanorods and surrounding director field **n(r)**. Green filled spheres at the ends of NP show boojums. (b) MSDs $\langle \Delta z^2 \rangle$ and $\langle \Delta x^2 \rangle$ respectively parallel and perpendicular to $\mathbf{n}_0$ in a planar cell ($d \approx 10$ $\mu$m). Solid lines are a linear fit to the data (open symbols). Inset shows a corresponding dark-field $8 \times 8$ $\mu\text{m}^2$ texture of GNR2. (c) Schematic diagram of GNR1 nanorod and disclination loop (a thick red line) around it. (d) MSDs $\langle \Delta z^2 \rangle$ and $\langle \Delta x^2 \rangle$, respectively, parallel and perpendicular to $\mathbf{n}_0$ vs $\tau$ for GNR1 in a planar cell ($d \approx 10$ $\mu$m).

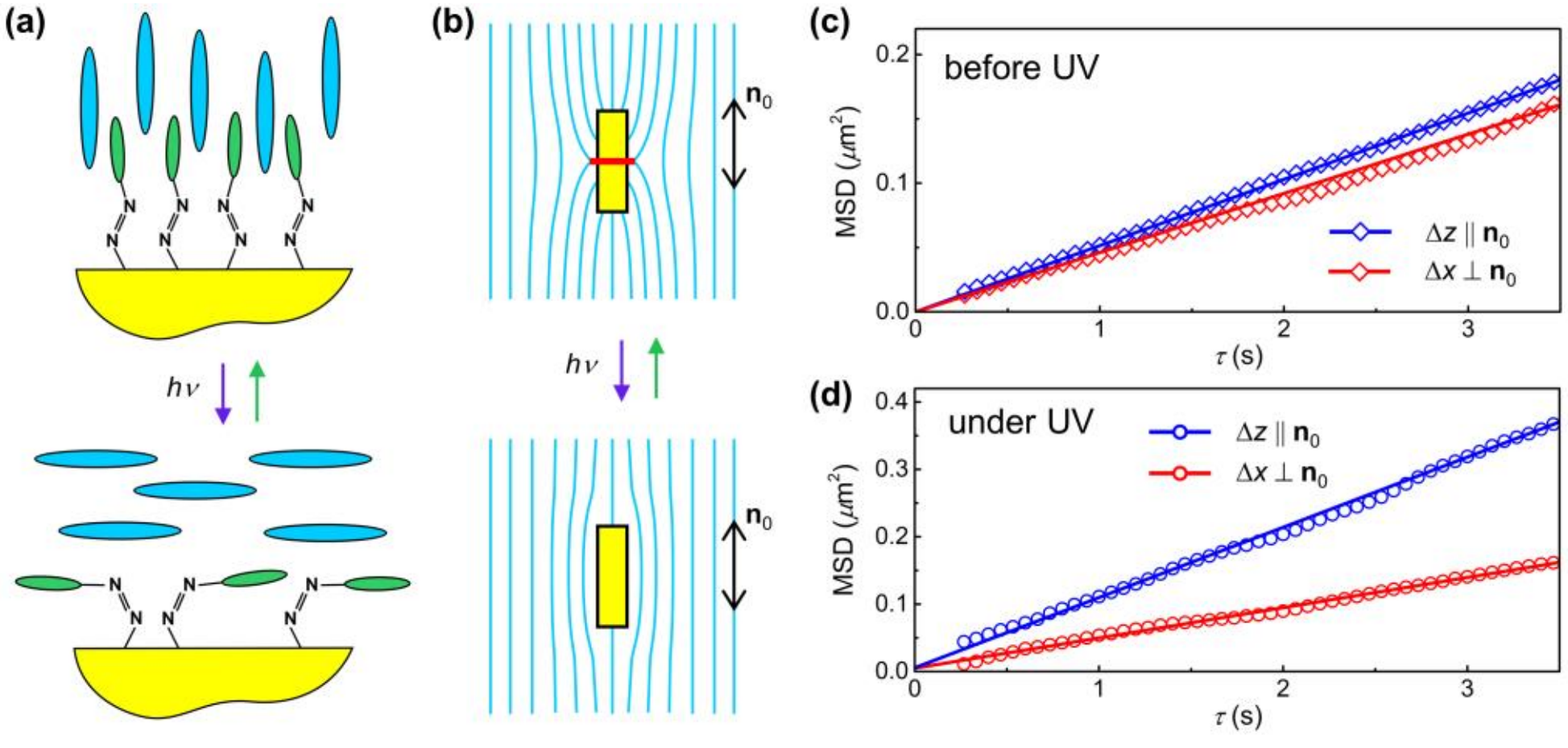

FIG. 5. (Color online) Diffusion of gold GNR4 nanorods with photosensitive capping in a nematic liquid crystal. (a) Schematic diagram showing reversible *trans-cis* isomerization of photosensitive capping molecules under the irradiation of UV light resulting in the change of the alignment of liquid crystals molecules (large blue ellipses). (b) Director field around the GNR4$_{trans}$ (top) and GNR4$_{cis}$ (bottom). (c), (d) MSDs $\langle \Delta z^2 \rangle$ parallel and $\langle \Delta x^2 \rangle$ perpendicular to $\mathbf{n}_0$ vs. $\tau$ for nanorods in a planar cell ($d \approx 10$ $\mu$m), respectively, before (c) and under (d) the UV light irradiation.

TABLE I. (Color online) Diffusion parameters of gold NPs of diameter $2R$ and length $L$ in nematic LCs in comparison to calculated for spheroids in isotropic liquid and simulated and experimentally measured for spherical colloids in nematics.

| | Nanoparticle and director fields | Dimensions (nm) | | Surface anchoring/ Director configuration | $D_a$ (μm²/s) | $D_b$ (μm²/s) | $D_a/D_b$ | $D_\parallel$ (μm²/s) | $D_\perp$ (μm²/s) | $D_\parallel/D_\perp$ |
|---|---|---|---|---|---|---|---|---|---|---|
| | | $2R$ | $L$ | | | | | | | |
| Anisotropic colloidal particle in anisotropic host, experimentally measured | CPN 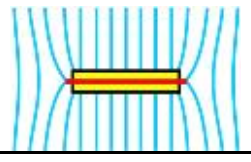 | 150 | 800 | Homeotropic/ Saturn ring | $0.59\times10^{-2}$ | $0.46\times10^{-2}$ | 1.28 | $(0.72\text{-}0.8)\times10^{-2}$[a] | $(0.72\text{-}0.88)\times10^{-2}$[a] | 0.82-1.11[a] |
| | CSN 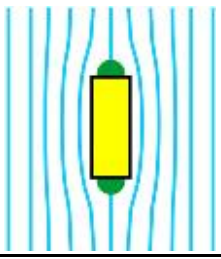 | 100 | ~500 | Tangential/ Bipolar | $1.11\times10^{-2}$ | $0.52\times10^{-2}$ | 2.14 | $1.11\times10^{-2}$ | $0.52\times10^{-2}$ | 2.14 |
| | NBN 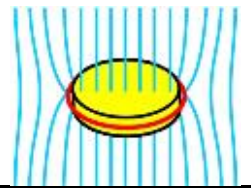 | ~500 | 100 | Homeotropic/ Saturn ring | $\sim6.9\times10^{-2}$ | $\sim2.6\times10^{-2}$ | ~2.59 | $\sim6.9\times10^{-2}$ | $\sim2.6\times10^{-2}$ | ~2.59 |
| | GNR1 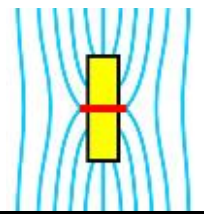 | 25 | 60 | Homeotropic/ Saturn ring | $4.12\times10^{-2}$ | $2.91\times10^{-2}$ | 1.42 | $4.12\times10^{-2}$ | $2.91\times10^{-2}$ | 1.42 |
| | GNR2 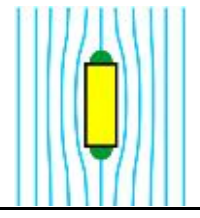 | ~12 | ~45 | Tangential/ Bipolar | $3.27\times10^{-2}$ | $2.05\times10^{-2}$ | 1.59 | $3.27\times10^{-2}$ | $2.05\times10^{-2}$ | 1.59 |
| | GNR3 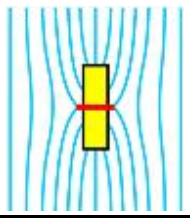 | ~12 | ~45 | Homeotropic/ Saturn ring | $2.92\times10^{-2}$ | $1.69\times10^{-2}$ | 1.73 | $2.92\times10^{-2}$ | $1.69\times10^{-2}$ | 1.73 |
| | $GNR4_{trans}$ 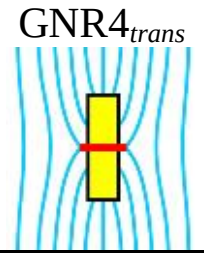 | ~20 | ~50 | Homeotropic/ Saturn ring | $2.57\times10^{-2}$ | $2.3\times10^{-2}$ | 1.12 | $2.57\times10^{-2}$ | $2.3\times10^{-2}$ | 1.12 |
| | $GNR4_{cis}$ 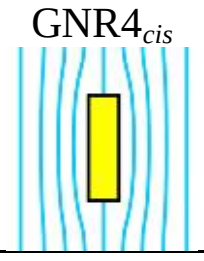 | ~20 | ~50 | Tangential/ Uniform | $5.2\times10^{-2}$ | $2.24\times10^{-2}$ | 2.32 | $5.2\times10^{-2}$ | $2.24\times10^{-2}$ | 2.32 |
| Anisotropic colloidal particle in isotropic host, calculated[b] | prolate spheroid 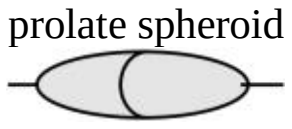 | 150 | 800 | - | $2.1\times10^{-2}$ | $1.6\times10^{-2}$ | 1.34 | - | - | - |
| | prolate spheroid | 100 | 500 | - | $3.3\times10^{-2}$ | $2.5\times10^{-2}$ | 1.33 | - | - | - |
| | oblate spheroid 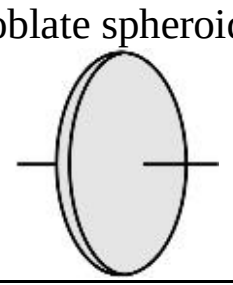 | 500 | 100 | - | $1.36\times10^{-2}$ | $1.78\times10^{-2}$ | 0.76 | - | - | - |
| | prolate spheroid | 25 | 60 | - | $18.2\times10^{-2}$ | $15.4\times10^{-2}$ | 1.18 | - | - | - |
| | prolate spheroid | 12 | 45 | - | $31.5\times10^{-2}$ | $24.7\times10^{-2}$ | 1.27 | - | - | - |
| | prolate spheroid | 20 | 50 | - | $22.4\times10^{-2}$ | $18.8\times10^{-2}$ | 1.19 | - | - | - |

TABLE I. (Continued.)

| | Nanoparticle and director fields | Dimensions (nm) | | Surface anchoring/ Director configuration | $D_a$ ($\mu m^2/s$) | $D_b$ ($\mu m^2/s$) | $D_a/D_b$ | $D_\parallel$ ($\mu m^2/s$) | $D_\perp$ ($\mu m^2/s$) | $D_\parallel/D_\perp$ |
|---|---|---|---|---|---|---|---|---|---|---|
| | | $2R$ | $L$ | | | | | | | |
| Isotropic colloidal particle in anisotropic host | Elastic dipole 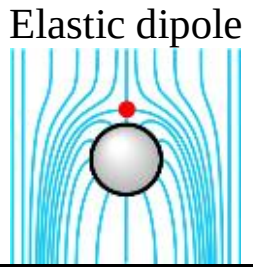 | 35 | 35 | Homeotropic/ Dipolar | - | - | - | $4\times10^{-2}$ [27][c] | $2.7\times10^{-2}$ [27][c] | 1.64 [39][d] 1.48 [27][c] |
| | Homeotropic elastic quadrupole 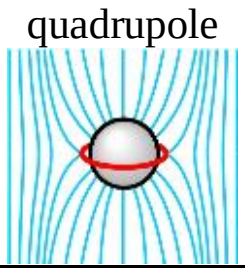 | 35 | 35 | Homeotropic/ Saturn ring | - | - | - | $5.6\times10^{-2}$ [27][c] | $3.67\times10^{-2}$ [27][c] | 1.72 [39][d] 1.53 [27][c] |
| | Planar elastic quadrupole 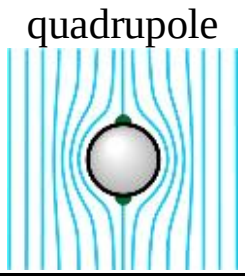 | 169 | 169 | Tangential/ Bipolar | - | - | - | - | $6.28\times10^{-2}$ [25][c] | 2.18-3.02 [33][d] 2.2-2.5 [63][c] |
| | Uniform 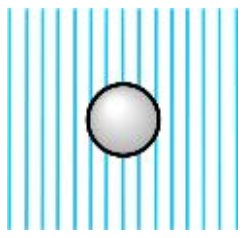 | 35 | 35 | Tilted/ Uniform | - | - | - | - | - | 2 [39][d] |

[a]Data for different orientations, see Fig. 1(c) and 1(d).

[b]Estimated using Eqs. (5) and (6).

[c]Experimental measurements.

[d]Numerical simulations.